# Review of Machine Learning Algorithms for Brain Stroke Diagnosis and Prognosis by EEG Analysis

Mohammad-Parsa Hosseini, Cecilia Hemingway, Jerard Madamba, Alexander McKee, Natalie Ploof, Jennifer Schuman, and Elliot Voss

*Abstract*— **Currently, strokes are the leading cause of adult disability in the United States. Traditional treatment and rehabilitation options such as physical therapy and tissue plasminogen activator are limited in their effectiveness and ability to restore mobility and function to the patient. As a result, there exists an opportunity to greatly improve the treatment for strokes. Machine learning, specifically techniques that utilize Brain-Computer Interfaces (BCIs) to help the patient either restore neurologic pathways or effectively communicate with an electronic prosthetic, show promising results when applied to both stroke diagnosis and rehabilitation. In this review, sources that design and implement BCIs for treatment of stroke patients are evaluated and categorized based on their successful applications for stroke diagnosis or stroke rehabilitation. The various machine learning techniques and algorithms that are addressed and combined with BCI technology show that the use of BCIs for stroke treatment is a promising and rapidly expanding field.**

## 1. Introduction

Strokes are the leading cause of adult disability and the fifth leading cause of death in the United States. These events occur when the blood supply to the brain is cut off, resulting in a lack of oxygen and nutrients to brain tissues. As a result, rapid necrosis and loss of brain function can occur within minutes and lasting deficits in motor control occur in over 85% of all cases.[1] With 800,000 cases of stroke each year, as well as the economic costs for both the patients and the healthcare system estimating about $23.6 billion,[95] there is a pressing need for improved prevention, diagnosis, treatment and rehabilitation protocols.[84,85]

Current treatment for strokes vary depending on the type of stroke: ischemic or hemorrhagic.[3] Ischemic strokes are the most common and occur when a clot blocks blood flow to the brain. It is treated either with medications to break up the clot or surgery to remove the clot. Hemorrhagic strokes occur when there is a bleed in the brain. Treatment involves controlling the bleeding and reducing pressure in the brain using medication, to counteract blood thinners, lower intracranial pressure or lower blood pressure, or surgery, to repair the damaged blood vessel. Once the initial problem is corrected, therapy depends on the long term effects the stroke had on the patient. Therapy resulting from stroke related paralysis is the same as therapy to treat any type or paralysis.

Recent research suggests that encephalography, or EEG, can potentially be used for diagnosis of acute ischemic attacks, monitoring for post-stroke epileptic events and outcome prognosis.[4] Foreman and Claassen have shown a correlation between cerebral blood flow (CBF) and changes in EEG signal. Specifically, during a stroke, there is a decrease in CBF. Additionally, Foreman and Claassen's analysis showed that as CBF decreases, the EEG signal will change to represent this loss of blood flow. With this information, EEG can potentially be used to detect, diagnose and monitor stroke progression. Real-time continuous monitoring could allow healthcare providers to assess damage and prevent secondary injury following stroke events to help mitigate symptoms and prevent future permanent damage.[4]

*Authors are with Santa Clara University, Department of Bioengineering, 500 El Camino Real, Santa Clara, CA 95053, mhosseini@scu.edu*

Machine learning, with its ability to quickly shift through, analyze, and classify data,[9] is a promising candidate for EEG analysis in stroke patients. Use of algorithms can lead to faster, more accessible and potentially accurate evaluation of patient data that will lead to more timely and effective care.

Currently, various machine learning algorithms are being explored for detection, classification, and characterization of stroke EEG signals. However, the most common application for machine learning in stroke analysis via EEG is rehabilitation, specifically through the use of brain computer interfaces (BCIs) [110]. Different types of BCI's have been used already for aiding in stroke rehabilitation and machine learning can be applied to improve the accuracy of the brain computer interface through reinforcement learning.[5]

This review aims to assess the current state of machine learning in stroke diagnosis and rehabilitation, specifically through the analysis of patient EEG signals. The paper will focus on the various machine learning algorithms, their overall accuracy and their various applications.

## 2. Review of Machine Learning

### 2.1 Machine Learning Overview

The first use of a neural network was in 1943, Neurophysiologist Warren McCulloch and mathematician Walter Pits published a paper about neurons and how they worked.[6] They created a model showing the use of an electrical circuit therefore, beginning the use of a neural network.[6] In 1950 then the first Turing Test was created by Alan Turing.[6] This test we see a to on the internet now and is fairly simple. It is used to make it so if a computer wanted to pass the test it has to convince a human that it is. Human too and not a computer.[6] In 1952, the first computer program was created that was able to learn as it ran and this is what we saw as the game checkers created by Arthur Samuel.[6] Not much else in the way of progression in machine learning was made until the 1980's and 1990's when neural networks started picking up popularity again.[5] Since then and now into the 21st century, machine learning is at the forefront of progression in the business and technology boom.[6]

Machine learning technology often gets confused and misidentified with Artificial Intelligence (AI).[7] Machine learning can be considered a part of the overall category of AI thus a branch of AI because most of what we use AI for is creating algorithms that are ultimately solved by some use of a machine learning technology but it can be so much more than that as well.[7,86] It is also used in large in the fields of computer science and engineering that needs to extract data based on particular pattern recognition.[86] This comes into play when a computer learns from a previously made mistake and after a repeated analysis of the data can master the tasks that had been previously too complex for the machine to process.[87] Machine learning is the ability to program computers in order to enhance the performance criterion of artificial intelligence by use of example data or past experiences.[9] It can also be thought of a technique that has the ability to automatically learn a model of a relationship between a set of descriptive features and target features from a given set of historical examples.[10] This can incorporate two different roles to optimize the performance criterion; the statistical role which receives the inference from a sample and a computer science role which uses an algorithm to solve an optimization problem and represents the evaluation of the model for inference.[9] For use in corporations and business, machine learning can be used to conduct predictive data analytics.[10] Predictive Data Analytics allows for the creation of business and data processes and computational models to assist in making data-driven decisions.[10] The real scale of how beneficial the use of machine learning comes into play when you want to build predictive models from large datasets with multiple given features.[10] For these large datasets to be run, algorithms created through machine learning techniques work by searching through possible prediction model data to determine the most capable model to show the relationship between the given descriptive features and a target feature.[10] The most popular of these algorithms used in machine learning are information based learning, similarity-based learning, probability-based learning and error based learning.[10]

There are four general applications for machine learning including association, supervised learning, unsupervised learning and reinforcement learning.[9] This section will focus on the difference between supervised and unsupervised learning. Supervised learning is when a particular person or group has an expert in that field that leads a participant through a specific test.[11] This type of learning has two types of techniques, continuous and discrete outputs.[11] Continuous outputs include predictions such in a regression style where discrete outputs are called classification which can classify data sets into different groups.[11] Unsupervised learning is done individually from an expert and finds the similarity and trends between different data sets.[11] Unsupervised learning can also be classified with two different techniques, continuous and discrete but in this case continuous refers to data reduction and discrete is called clustering.[11]

Every Machine Learning model must be comprised of four main ingredients before you can even begin to run your algorithm.[7] These ingredients are the data, the model, the objective function, and the limitation algorithm that you are wanting to run.[7] The data that will be used is usually historical data which is readily available to the public but sometimes you must incorporate your own research to create a data set that is particular to your specific research.[7] The next ingredient was the model which is the part of the machine learning process that you can train to use the dataset in your specific research.[7] The third ingredient is the objective function which comes into play for when you are ready to have an output given to you by your model that is as close to reality as possible.[7] The objective function is used to determine the accuracy of the output that your model has provided.[7] The final ingredient of a machine learning model is the optimization algorithm which is the mechanics that is used to vary the parameters of the model to optimize the objective function and achieve the proper output.[7] These ingredients are more than just steps, machine learning is iterative.[7] The machine learning process does not stop after one rotation through the ingredients, the process continues until you have varied your parameters to where you can no longer optimize the model any further.[7]

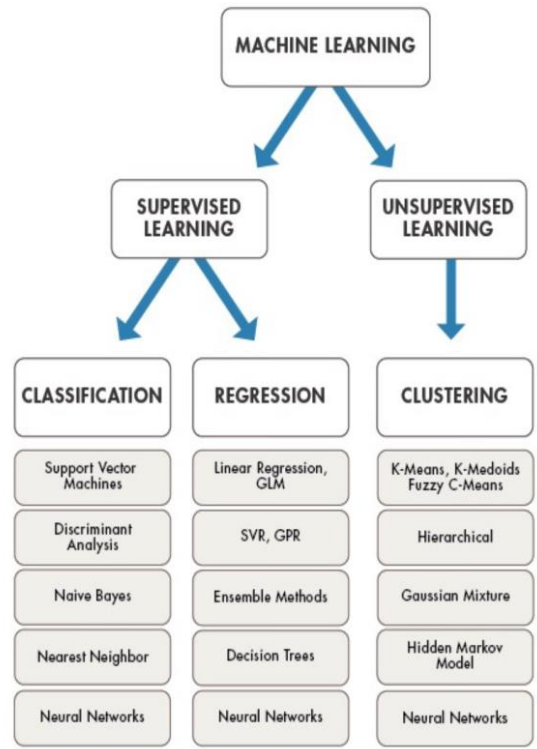

*Figure 1: Machine Learning algorithms split into supervised learning categories which includes the specific algorithms used in this classification and the unsupervised learning category which includes the specific algorithms used in this classification (Bhattacharjee 2017).*

Machine learning has become an important tool in developing systems to interpret neuroimaging datasets that can provide valuable information for further research into the interactions, structures and mechanisms within the brain and the behavior that can be involved with certain neurological disorders.[88,89] These systems of machine learning are now being implemented in clinical neuroscience research for the development of imaging-based diagnostic and classification systems of the brain's neoplasm.[90] It can also be used in diagnosing psychiatric disorders,[91] epilepsy,[92] neurodegenerative disorders,[93] and demyelinating disorders.[94]

With any computer and algorithm-based technology, problems can always arise. Machine learning can cause datasets or algorithms to become more generalized and not specific enough for different research topics.[10] The generalization can be caused due to a bias in the average model compared to the true model which can then create an error due to inaccurate assumptions made due to how the model looks.[12] Generalization can also be measured by variance and how much the model's estimates are different from other training sets.[12] The wrong inductive bias can also be chosen which can then lead to under and overfitting of the data.[8] Thus leading to creating the right balance between the complexity of the model and the simplicity of it being the most challenging part in machine learning.[10]

*2.2 Brain-Computer Interfaces*

As early as 1973, Vidal questioned whether there was potential for electrical brain signals to convey information which produce a command for devices such as prosthetics. This is when the use of brain signals were being used to control a prosthetic arm.[13] He created the Brain-Computer Interface Project which attempted to answer his question of whether brain signals could in fact control an external device. This early research by Vidal became the foundation to subsequent research on brain computer interfaces (BCIs).[14] Brain-computer interfaces (BCI) systems were now able to be used to record the brain activity that could be used to communicate between a human's brain and a computer in ways that can control the environment in a way that can be compatible with the intentions of what humans would want.[15]

Initially, research on BCIs progressed slowly. However, small discoveries regarding brain signals ability to control external devices began to emerge. By 1980s, Elbert et al. demonstrated that individuals could control the direction of a rocket image through biofeedback of EEG activity.[16] P300 potentials occur in the brain when an individual recognizes, decides, or categorizes visual stimuli.[17] This concept was used by Farwell and Donchin, when they had the participants in their study use their P300 event-related potentials to spell words on a computer screen.[18]

BCIs rely on gathering data from electroencephalographic signals from the brain and extracting features from them through signal processing and other feature analysis.[19] Subsequently, by running the extracted data through a predetermined machine learning model, the input data can be interpreted as a computer-generated output which prompts the facilitation of a desired action. Thus, the BCI system can be broken down into four components: signal acquisition, feature extraction, feature translation, and device output.[18] Signal acquisition refers to the gathering and measuring of electroencephalographic signals through a sensor such as an EEG.[19] These signals are magnified in order to process them electronically, and electrical noise or other unwanted features of the signal are filtered out. Once this signal is modified and transmitted into a computer, the signal is analyzed to distinguish important characteristics through a process called feature extraction. Commonly extracted features of EEG activity may include time-triggered EEG response amplitude and latencies, such as P300 waves,[20] or power within EEG frequency bands, such as those from sensorimotor rhythms.[21] Removal of environmental and physiologic artifacts found within the signal like electromyographic signals are removed to increase accurate measurement of significant brain signal features and reduce confounding variables. These extracted signal features serve to decipher the user's intent. The extracted data will undergo feature translation with the use of a machine learning model algorithm to convert the features into a command desired by the user. Ideally, the machine learning model would be able to adapt and accommodate for dynamic learned changes or spontaneous changes to the signal features for increased accuracy.[22] The signal processing from feature extraction and feature translation will contribute to a device output. The possibilities of output are immense from letter selection[16,26] and cursor movement on a computer,[23,24,26] driving a wheelchair,[25,26] to manipulating a robotic arm.[19,26]

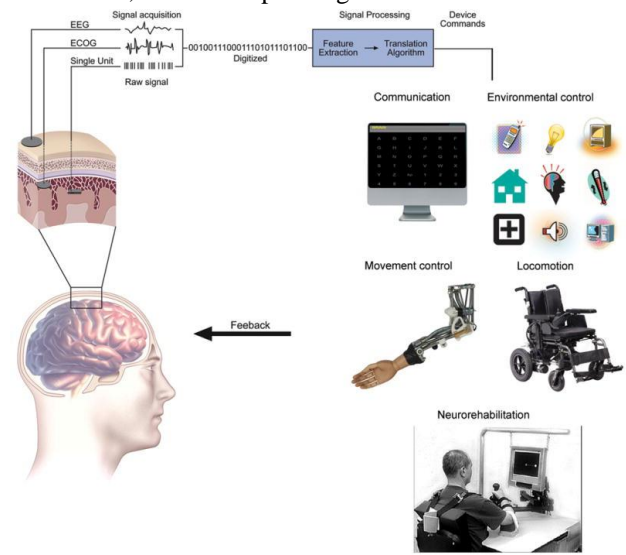

*Figure 2: Key aspects of the BCI system.[65,66] Intent for action is carried from the user by electrodes detecting EEG readings. The readings undergo feature extraction and are translated in commands. Those commands are used to activate the desired technology.[67] Figure designed by Mak, J. N., & Wolpaw, J. R. (2009).*

Since 2000, BCI research is growing at a exceedingly faster rate due to the technology now available to researchers. Current literature on BCI interventions have used participants with stroke as well as on healthy participants as a proof of concept. or as a control variable.[27] Although in the minority it is expanded that event-related desynchronization (ERD) or event-related synchronization (ERS) occurs in healthy patients as well as stroke patients.[28]

Most studies use a combination of stroke patients and healthy participant controls in order to show the effectiveness of the method in different populations.[29]

Current research strives to develop BCI systems in order to be used in the medical setting. With BCI technology, an individual who has weakened or completely lost voluntary movement may be able to regain functionality through using brain signals.[30] Most studies in the current literature investigating the implementation of the BCI system on stroke patients focus on the rehabilitation of the upper body, specifically muscles from the shoulder to hand. However, BCIs can theoretically be used to facilitate movement of any muscle with voluntary motor impairment.

There are two types of interfaces currently used in stroke rehabilitation: invasive and non-invasive.[13] Invasive techniques involve putting an implant within the patient's brain[35] so that electrodes can record action potentials (APs) and local field potentials (LFPs). On the other hand, non-invasive BCI's use the individual's EEG readings. The invasive BCI's have been shown to have a more stable, more accurate performance, but is less accepted as a practice because of the additional risks.[13] Currently, there does not seem to be any invasive commercial BCI's available, although there are some non-invasive options. Non-invasive options suffer from a loss of accuracy because the EEG has to pick up signals farther away, through tissue and bone. Artifacts can be reduced through filtering, however, the signal is not as stable as the invasive versions.[13]

With any technology, errors can arise and there must be ways to test for those errors and correct them. With BCI errors and error-related potential (ErrP) has been used as the ERP compote that can correct the BCI errors.[62] An ErrP can be used when a mismatch is found between the subject's intentions to perform a wanted task and the response that is received by the BCI.[63] This can then be used to allow for the user to control the behavior of an external autonomous system and be able to correct it in real time.[64]

## 3. Review of Machine Learning in Stroke

### 3.1 Diagnosis

The implementation of machine learning in the diagnosis of stroke has helped increase the amount of early identification of imaging diagnostic findings,[96] ability of estimating the time of onset,[97] the lesion segmentation,[98] the fate of salvageable tissue,[99] and finally the outcome of the patient and the possibly long-term issues they may have to endure.[100]

### 3.1.1 Neural networks

A convolutional neural network (CNN) has been previously used to distinguish between control and stroke EEG data, which is more readily available than CT scans.[31] Applying early stopping and batch normalization techniques accelerated the group's classification model. F scores using the CNN are higher than that of the Naive Bayes model, and was shown to achieve better results than the compared classifiers, and is able to outscore neural networks with a lower number of epochs.[31] This CNN has a stochastic aspect compared to the Naive Bayes model.[31] For this model, classifying the normal category is easier than identifying the stroke category, and a 100% prediction success rate for the normal category is achieved by this CNN.[31] Difficulties arise when classifying stroke patients due to differing stage levels of stroke across the stroke patients, which limits the accuracy of classifying in the stroke category.[31]

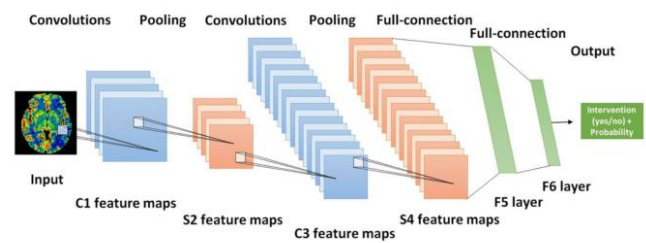

*Figure 3: A convolutional neural network deep learning map showing steps from the input to the output.*[34]

CNNs have outperformed baseline learning methods and linear classifiers when discriminating normal from pathological EEG.[33] Evaluation of accuracies has shown that with reduced training-time, Deep CNN accuracy decreased, with the highest achievable accuracy after one minute.[33] Construction of a deep CNN to evaluate EEG signals for pathology could assist in stroke diagnosis when binary[101-103], but more fine-tuned diagnoses are still limited.[33] Real time stroke diagnosis may be assisted by improved deep CNNs that can enable for better recognition of physiological and pathological change in EEG[33, 104,105,106,107].

Neural networks able to perform automatic feature extraction and feature selection circumvents limitations to manual features, as described by Cheng, et al.[34] This is applicable in BCI applications for stroke recognition.[34] EEG data is bandpass filtered into sub-bands by a sliding window strategy.[34] Different spatial-spectral features are then extracted and fed into a deep neural network to reveal spatial and spectral patterns.[34] This method has higher classification accuracy than several other state of-the-art approaches.[34]

*3.1.2 Support Vector Machine*

As previous research has concluded, EEG is a potentially powerful tool for diagnosing stroke; however feature extraction from these signals can be quite complex.[4,37] SVM, which uses kernels to map samples from one feature space to another, has been shown to be exceptionally powerful in pattern recognition for higher dimensional and nonlinear problems.[38] Although EEG signals can be difficult to interpret, this nonlinear method of classification has shown to be especially useful for mental task recognition.[68] Previous research has looked at directly comparing linear and nonlinear methods for the specific task of classifying mental tasks. A 2003 study looked at linear discriminant analysis (LDA) compared to neural networks (NN) and support vector machines (SVM).[73] Overall, while LDA is a powerful method for classification, it is limited to linear analysis.[74] Additionally, neural networks, while flexible, require large sets of data.[75] Therefore, SVM is a strong candidate for nonlinear analysis of small data sets. However, while SVM is a promising method for accurate classification of stroke related EEGs, it can be difficult to select the correct kernel for a given task. Therefore, the use of multiple kernels has been proposed.

Multiple kernel learning (MKL) SVM relies on a series of predefined kernel and determines a nonlinear combination. A 2008 paper by Rakotomamonjy et al. proposed a MKL-SVM method aimed at guiding the iterative process involved in this algorithm called SimpleMKL. SimpleMKL uses a weighted 2-norm regularization formulation while constraining the weights.[77] This method was shown to be applicable for problems involving regression, clustering (one-class classification) or multiclass classification.[77]

Li et al. explored the use of a three stage process that explores the use of a multiple kernel learning SVM (MKL-SVM) for classification of cognitive task EEG to be applied to BCIs [108,109]. First, raw EEG signals were pre-processed to improve the general quality of the signal. Next, features were extracted via wavelet packet entropy (WPE) and Granger causality flow. Next, Li et al. employed the SimpleMKL proposed by Rakotomamonjy et al. which relied on use of linear combination of multiple kernels for performance of 2-class, 3-class, 4-class, and 5-class recognition. This method accounts for the given weights of each kernel to ensure a more accurate algorithm. Results showed that the MKL-SVM outperformed the single kernel SVM approach, achieving 99.2% accuracy for 2-class classification and greater than 75% for multiclass classification.[38]

While Li et al. focused on improvement of the SVM classification algorithm, Liu et al.[39] proposed an improved method for feature extraction, an essential component to SVM success. In there paper, Liu et al.[38] developed a common spatial-spectral boosted pattern (CSSBP) for enhanced feature extraction from EEG signals obtained from stroke patients. The success of common spatial patterns (CSP) relies heavily on predetermined spatial-spectral conditions for filtering; however, it is often difficult to identify these settings for stroke patients. Therefore, Lui et al. propose a method that combines CSP with boosting strategies. Upon comparison to seven different state-of-the-art feature extraction algorithms including power spectral density, phase synchrony rate, the original CSP, regularized CSP (RCSP), the sub-band CSP (SBCSP), CSSP and CSSSP. Following feature extraction, a linear SVM was used for classification and the results showed that the combined CSSBP and SVM method outperformed all other methods and was able to achieve 70% accuracy following a two-month rehabilitation period.[39]

*3.1.3 Other ML methods for stroke diagnosis*

In stroke diagnosis, it is important to quantify the degree of movement impairment in an individual as this has a strong impact on the individual's resulting quality of life. One way to do this is by using event-related synchronization (ERS) as an assessment feature for the degree of movement impairment in a stroke patient.[40] In Kim et al.[40], the correlation between presence of ERS after hand movement between stroke patients and healthy controls was investigated. It was found that there was a statistically significant difference in the power of ERS and the occurrence timing of ERS between these two groups. For the chronic stroke patients, the power of ERS was lower and the timing was delayed. A cortical activation model then demonstrated that this delay and low power of ERS can be interpreted as, the patient group had to put more effort in to executing the hand movement as ERS indicates a decrease in cortical activation. Additionally, ERS was again observed between stroke patients with mild motor impairment EEG and severe motor impairment and the same trend was observed. As a result, Kim et al., demonstrates that ERS can be used as an assessment feature when diagnosing the degree of motor impairment of stroke patients, further assisting in stroke diagnosis and classification through the use of EEG signals.

Additionally, stroke diagnosis and accurate classification using EEG signals is important for its potential application in BCI-FES (Brain-computer interface - functional electrical stimulation) stroke rehabilitation systems. In these systems, EEG recognition techniques are combined with FES in order to help patients reconstruct neural circuits between paralyzed limbs and corresponding brain areas that become damaged and destroyed after a stroke. Currently, a common shortcoming exists when using EEG data for these systems, as they use conventional methods for classification and feature extraction.[41] Methods such as common spatial patterns (CSP) are known to have poor discriminant abilities. In Zhang et al.[41], to avoid this problem, they visualize the original spatial patterns and then build a Gaussian Mixture Model (GMM) to represent this. In this way, GMM is used for advanced feature learning, as it is a probabilistic model that uses Gaussian distribution to represent the presence of sub-populations within a larger population. In this paper, they applied the model both as a filter in a pre-processing module and as a classifier in a classification module. When compared to CSP-SVM, GMM as a filter and as a classifier outperforms in terms of accuracy, achieving a high accuracy of 77% in the case of one of the test subjects. As a result, Zhang et al., demonstrates that GMM is yet another viable and more accurate way to filter and classify EEG data for diagnosis of stroke patients.

Another method used in machine learning to help with stroke diagnosis is sensorimotor rhythms (SMR) paradigms. The use of a sensorimotor rhythms paradigm is now one of the most popular motor imagery paradigms.[69,70] In an SMR paradigm, the movement that is being imagined is defined by the imagination of the kinesthetic movements of the large body parts such as our limbs that could be read as modulations of brain activity.[71] This paradigm has also been used in testing between healthy patients and patients that have suffered strokes.[72]

### 3.2 Rehabilitation

The process of brain reorganization in patients that exhibit chronic stroke symptoms usually results in the over-use of the contralesional hemisphere as well as the under-use of the lesioned hemisphere.[52] This will lead to the increased inhibitory activity in the contralesional area to the ipsilesional hemisphere.[52] This increased inhibitory influence will end up blocking out the excitatory reorganization of the remaining healthy and intact areas around the lesion and will slow the recovery of the affected motor system.[76] The following methods and strategies discussed rely largely on manipulating neural circuits in order to regain the motor recovery lost by the side effect of stroke.[78]

### 3.2.1 Rehabilitation Methods

In terms of rehabilitation for stroke patients that have any category of paralysis, use of brain-machine interfaces (BMI) is proving to be one of the only therapeutic options in helping the patient regain their loss of motor function.[42] Brain-machine interfaces (BMI) is a system used to record, decode and translate the brain signals into a useful action or behavior without having to involve a total motor system.[52] The use of BMI systems have been increasing over the last couple decades and have been developed for use in communications, control of different devices and for total rehabilitation.[79,80,81,82] Use of BMI in motor loss works towards restructuring the motor pathways by producing a contingent link between the damaged brain section causing the paralysis and the peripheral nerves and muscles needed to regain the movement.[43] To create this link between the brain and the nerves and muscles, brain activity must be monitored through the use of electroencephalography (EEG).[42] The use of wireless EEG's in this method has begun to have a major importance because they simplify the overall set up and reduce any error in the movements of the wires that may generate artifacts in the signals.[83] In López-Larraz et. al, the goal was to test an EEG-EMG hybrid Brain-Computer Interface (BCI) to improve the accuracy of sensing for stroke victims with complete hand paralysis.[42] Within their research, they tested 20 chronic stroke patients that produced uncontrolled compensatory activities during movement that masked relevant brain activation, ultimately making it difficult to interpret the EEG signals.[42] Along with EEG signals, it was shown that residual EMG activity in patients with complete hand paralysis can be classified for intention detection.[44] They tested EEG-only, EMG-only, and EEG-EMG combined. The EEG recorded with 16 electrode Acticap system (Fp1, Fp2, F3, Fz, F4, T7, C3, Cz, C4, T8, CP3, CP4, P3, Pz, P4, and Oz with grounds AFz and FCz).[42] They used a block-based N-fold cross-validation method for training, where N was the number of blocks performed by each participant.[42] One block was used for testing and the remaining were used for training to create a classifier for each movement intention.[42] For EEG, the average power for alpha and beta were calculated during feature extraction.[42] They found throughout this study that the hybrid system reduced false positives and enhanced true positives.[42]

Another method of rehabilitation for stroke patients that have any category of paralysis is the use of Motor Imagery (MI), which is the mental process of imagining movements without the actual physical movement occurring.[45] By performing MI, this activates the brain regions in the sensorimotor network that creates a stimulation similar to the actual physical movement occurring.[46] The purpose of Ang's paper is to review the three strategies of using Brain-Computer Interface (BCI) to detect Motor Imagery using EEG data.[47] The three strategies discussed are as follows: an operant condition that employed a fixed model, machine learning that employed a subject-specific model computer from calibration, and an adaptive strategy that continuously computes the subject-specific model.[47] Within the review, it was found that most papers used the operant conditioning as a control and not many applied it to the rehabilitation process because it requires the subject to undergo multiple sessions of learning in order to control a specific EEG rhythm needed.[47] In the second strategy discussed, the subject only has to undergo a single calibration session.[47] In this strategy though, there is a non-stationarity in the session-to-session transfer between the calibration session and the online feedback sessions.[47] As a result, the third strategy which is adaptive is reviewed.[47] Using the adaptable strategy, on averages, a 12 percent accuracy improvement was observed.[47]

The use of EEG's is a great technology to study the real-time brain activity during the entire BCI therapy session with high temporal resolution but neuroimaging methods have given us the ability to study both the large-sale and the small-scale reorganization of brain networks at a somewhat higher spatial resolution.[48] The use of interventional therapy and the incorporation of brain-cover interface (BCI) technology is also showing as a promising method for recovering motor movement in stroke survivors.[48] Unfortunately we do not have enough of an understanding of this technology to be able to incorporate it with functional networks outside of a motor network.[48] Until we have more information about how this technology can be use with other networks, Mohanty's paper researched the impacts of BCI-therapy on resting-state functional connectivity in stroke patients.[48] Their method used machine learning models to look for classifieds that would be able to separate users into before, during, and after BCI therapy states.[48] Using a permuted-block design the characteristics of gender, stroke chronic it's, and severity of motor impairment was used to separate the participants into one of two groups, either crossover control group or the BCI therapy group.[48] After conducting the research, there were changes found in the front-parietal task control, the default mode, the subcortical, and visual region which further indicates that more than just motor function can be improved in stroke patients using BCI-mediated rehabilitation.[48]

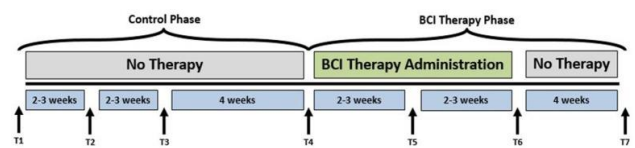

*Figure 4: Study paradigm. This figure shows the points where neuroimaging data was collected. This is represented by: T1: control baseline 1, T2: controls baseline 2, T3: control baseline 3, T4: therapy baseline, T5: mid-therapy, T6: post-therapy, and T7: 1-month post-therapy. The crossover control group completed visits T1 through T7. The BCI therapy group completed visits T4 through T7 only.*

The paralysis that occurs after the event of a stroke or neurotransmitter is among the leading causes off long term disabilities in adults.[43] Even without having true motor movement in their limbs, most stroke patients are still able to imagine what the movement would feel like in their limbs thus allowing for the brain to fire he correct neurons that would initially be used to move the desired limb.[43] Evaluations of efficacy contributed with daily brain-machine-interface training has also been done to determine if the hypothesized beneficial effect of physiotherapy in patients that have had severe motor movement problems increased.[43] This was done through the use of a double blind sham controlled design proof of concept study.[43] The physiotherapy was used following the BMI or placebo sessions during the test on stroke and paralysis patients.[43] An EEG signal was used to track sensorimotor rhythm on paralyzed limbs to detect intentions that the test subject had to move that given limb.[43] As found in a previous study, the learning to control desynchronization of ipsilesional sensorimotor rolandic brain oscillations (SMR) after a stroke has occurred can be translated into small grasping movements of an orthosis that is attached to a paralyzed limb.[43] In this study, they saw a movement of orthosis if the EEG signal remained continuous in the motor intention classification region, if not then the ground state was retained.[43] This research was done without a true machine learning algorithm but did show the potential application for BMI development for use in recognition of intention to move post-stroke via BMI.[43]

It has also been proven that stroke and accompanying neuroplastic changes that are correlated with the rehabilitation process for after-stroke patients can led to significant inter-subject changes within the EEG features that are suitable for mapping as a part of the neurofeedback therapy.[49] This would include individuals that had also scored largely similar with current conventional behavioral measures.[49] The stroke-affected EEG datasets showed a lower 10-fold cross-validation results than the comparably healthy EEG datasets.[49] Therefore, it can be believed that by motor retraining with the use of BCI, we could initially tailor this to the individual patient's needs.[49] This was found by using a 32-channel EEG and recording the finger-tapping task from 10 healthy subjects for an entire session and 5 stroke patients from two sessions that were approximately taken 6 months apart.[49] A aching learning model was created using an off-line BCI design that was based on a Filter Bank Common Spatial Pattern (FBCSP) that has the adaptability to test and compare the efficacy and accuracy of the training of a rehabilitative BCI with both the stroke affected patients and the healthy patients.[49]

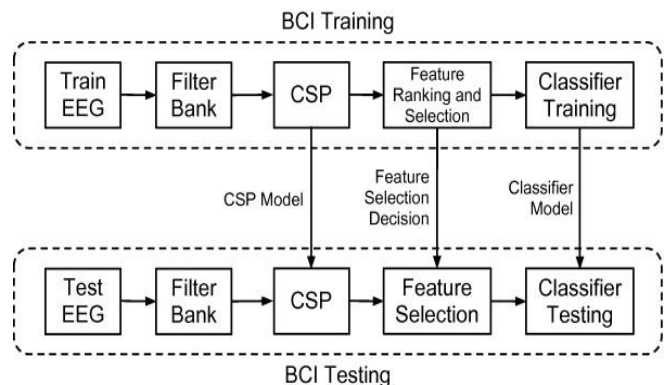

*Figure 5: Simplified diagram of the off-line Brain-Computer Interface implementation used (Leamy 2014).*

*3.2.2 Implementation*

Implementation of the BCI system to patients can have a positive effect in patient recovery. The BCI controls a system that helps the patient move in natural ways, despite the patient being unable to move on their own. The two main forms of implementation involve robotic limbs and functional electrical stimulation (FES).[50]

Robotics limbs have been around since the first case in the 90s where engineers from MIT used their 5 degree of freedom robotic workstation to create motion in the hand of the patient.[35] No study was officially run in this paper, but the possibility is discussed. Robotic limbs have expanded into multiple different fields of rehabilitation ranging from hand motion to gait correction.[36]

In one study, the robotic limb provided variable amounts of resistance and assistance for movement which was dependent on the strength of the individual's hands and the amount of resistance the participant was able to move against on their own. By using robotic limbs to support movement as needed, there is potential for future application to limb rehabilitation for cases in which an individual has positive prognosis for regaining movement and control of their affected limb.[51] FES is a treatment that uses small electrodes which are placed on the patient's muscles to stimulate the muscle to move. This is done with paralyzed or weakened muscles that can no longer naturally move to their full extent. Electrodes are attached along the patient's arm in order to stimulate the muscle to perform as it would if was fully operational.[52] FES can be used in an array of applications, from movement of appendages to more internal functions like swallowing and breathing.[53, 54]

Supplementary Table 1: Review of Machine Learning Algorithms for use in Stroke Disease Lifetime

| Primary Source | Machine Learning Algorithm | Findings or Application | Reference |
| --- | --- | --- | --- |
| **Giri, et al. (2016)** | 1DCNN | F-Score 0.861, precision 0.870 | [31] |
| **Schirrmeister, et al. (2017)** | Convolutional NN | Accuracy ≈85% (vs. ≈79%) | [33] |
| **Cheng, et al. (2018)** | Deep NN | Mean 1 Accuracy = 71.5%, Mean 2 Accuracy = 67.7% | [34] |
| **Li, et al. (2014)** | MKL-SVM | 99.2% Accuracy for 2-class classification; 75% Accuracy for multiclass classification | [38] |
| **Lui, et al. (2016)** | CSSBP with SVM | 70% Accuracy | [39] |
| **Kim, et al. (2016)** | Independent Component Analysis | ERS in the beta band (13 - 32 Hz) only present for healthy patients | [40] |
| **Zhang, et al. (2013)** | Gaussian Mixture Model | High Accuracy = 77% | [41] |
| **Lopez-Larraz, et al. (2018)** | Block-based N-Fold cross-validation | Hybrid system reduced false positive and enhance true negative | [42] |
| **Ang, et al. (2017)** | Operant conditioning; subject-specific model; adaptive strategy | 12% motor imagery Accuracy improvement using adaptive strategy | [47] |
| **Qiu, et al. (2017)** | Support Vector Machine (c) | Motor Imagery accuracy for paretic hands was not substantially worse than with non-paretic hands | [57] |
| **Bhagat, et al. (2016)** | binary SVM with adaptive window technique | True Positive Rate = 67.1 ± 14.6% after 5 days | [58] |
| **Gatti, et al (2018)** | Convolutional NN | 80% for healthy patients, 60% stroke patients | [59] |
| **Suwannarat, A., Pan-ngum, S., & Israsena, P. (2018).** | Compared LDA and SVM | Difference between the two is not statistically significant | [27] |

| Mohanty, et al. (2018) | Support Vector Regression (SVR) | Linear and nonlinear SVR models were compared and found that they performed similarly with nonlinear being slightly more generalizable | [48] |
|---|---|---|---|
| López-Larraz, et al. (2017) | Linear regression artifact removal method | Demonstrates the importance of filtering data to maximize efficacy | [50] |

There are other, less common forms of treatment like using monitors or virtual reality which capitalizes on the use of the MI rehabilitation method, in order to facilitate visualization of the patient moving his/her appendage. For example, when the patient imagines flexing with their fingers, the imagine located above their hand, shown on a screen will change to look as though it is being manipulated by their intention to move.[55]

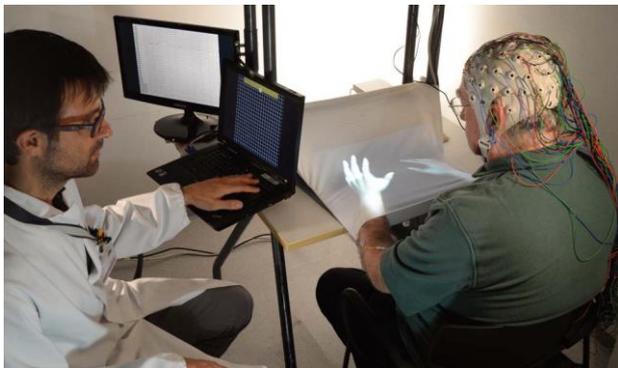

*Figure 6: Patient is seated with hand under a screen which projects visual representations of the patient's hands in motion. Figure taken by Ron-Angevin, R., & Díaz-Estrella, A. (2009).*

Implementation of BCIs is hindered by the amount of artifacts in EEG data. Non-invasive BCI methods like using EEG data is noisy and can cause misreading from the BCI. One study attempted to filter already taken EEG readings in order to create a better system to no avail.[56] This paper did not find any improvement from filtering the data while another paper found similar averages around 60-70% accurate.[57]

One study was done on the difference between training of the paretic hand relative to the non-paretic hand.[57] The researchers investigated whether neurological damage caused during a stroke could inhibit a patient from being able to "imagine" moving their hand. The article found that there was no substantial difference between training the paretic hand vs the non-paretic hand.[57] This shows that at least in some cases, damage from stroke does not stop a patient's ability to imagine movement.

## 4. Conclusion

Non-invasive BCI technology for application in stroke diagnosis and rehabilitation is a rapidly expanding and improving field. Various combinations of feature extraction and classification methods for machine learning show increasing accuracy when it comes to proper stroke diagnosis and rehabilitation when paired with BCI technology. The current reviewed evaluated these EEG-based BCI paradigms according to the classification of application for stroke diagnosis or stroke rehabilitation. Within the diagnosis classification, machine learning techniques such as convolutional neural networks, support vector machines, gaussian mixture models, etc. were reviewed. For stroke rehabilitation, the varying techniques were analyzed and categorized based on their data acquisition and analysis methods or the method of implementation. Overall, this review demonstrates that the use of BCI technology for stroke treatment, when paired with various combinations of machine learning techniques, is a promising field that shows improved accuracy in both stroke classification and patient intention differentiation when applied to stroke diagnosis and stroke rehabilitation, respectively.

___________________________________________

> REPLACE THIS LINE WITH YOUR PAPER IDENTIFICATION NUMBER (DOUBLE-CLICK HERE TO EDIT) <    14with minimal subject training. *Journal of neuroengineering and rehabilitation*, 6(1), 14.

[26] McFarland, D. J., Krusienski, D. J., Sarnacki, W. A., & Wolpaw, J. R. (2008). Emulation of computer mouse control with a noninvasive brain-computer interface. *Journal of neural engineering*, 5(2), 101-10.

[27] Rebsamen, B., Guan, C., Zhang, H., Wang, C., Teo, C., Ang, M. H., & Burdet, E. (2010). A brain controlled wheelchair to navigate in familiar environments. *IEEE Transactions on Neural Systems and Rehabilitation Engineering*, 18(6), 590-598.

[28] Mak, J. N., & Wolpaw, J. R. (2009). Clinical Applications of Brain-Computer Interfaces: Current State and Future Prospects. *IEEE reviews in biomedical engineering*, 2, 187-199.

[29] Suwannarat, A., Pan-ngum, S., & Israsena, P. (2018). Comparison of EEG measurement of upper limb movement in motor imagery training system. *Biomedical engineering online*, 17(1), 103.

[30] Pfurtscheller, G., & Da Silva, F. L. (1999). Event-related EEG/MEG synchronization and desynchronization: basic principles. *Clinical neurophysiology*, 110(11), 1842-1857.

[31] Mohanty, R., Sinha, A. M., Remsik, A. B., Dodd, K. C., Young, B. M., Jacobson, T., ... & Kang, T. J. (2018). Early Findings on Functional Connectivity Correlates of Behavioral Outcomes of Brain-Computer Interface Stroke Rehabilitation Using Machine Learning. *Frontiers in neuroscience,* 12.

[32] Shih, J. J., Krusienski, D. J., & Wolpaw, J. R. (2012, March). Brain-computer interfaces in medicine. *In Mayo Clinic Proceedings* (Vol. 87, No. 3, pp. 268-279). Elsevier.

[33] Giri, Endang Purnama, Mohamad Ivan Fanany, Aniati Murni Arymurthy, and Sastra Kusuma Wijaya." Ischemic stroke identification based on EEG and EOG using ID convolutional neural network and batch normalization." In 2016 International Conference on Advanced Computer Science and Information Systems (ICACSIS), pp. 484-491. IEEE, 2016.

[34] Feng, R., Badgeley, M., Mocco, J., & Oermann, E. K. (2018). Deep learning guided stroke management: a review of clinical applications. *Journal of neurointerventional surgery*, 10(4), 358-362.

[35] Schirrmeister, R., Gemein, L., Eggensperger, K., Hutter, F., & Ball, T. (2017, December). Deep learning with convolutional neural networks for decoding and visualization of EEG pathology. In *2017 IEEE Signal Processing in Medicine and Biology Symposium (SPMB)* d(pp. 1-7). IEEE.

[36] Cheng, D., Liu, Y., & Zhang, L. (2018, April). Exploring Motor Imagery Eeg Patterns for Stroke Patients with Deep Neural Networks. In *2018 IEEE International Conference on Acoustics, Speech and Signal Processing (ICASSP)* (pp. 2561-2565). IEEE. https://ieeexplore.ieee.org/document/8461525

[37] Hogan, N., Krebs, H. I., Charnnarong, J., Srikrishna, P., & Sharon, A. (1992, September). MIT-MANUS: a workstation for manual therapy and training. I. In [1992] *Proceedings IEEE International Workshop on Robot and Human Communication* (pp. 161-165). IEEE.

[38] Do, A. H., Wang, P. T., King, C. E., Chun, S. N., & Nenadic, Z. (2013). Brain-computer interface controlled robotic gait orthosis. Journal of neuroengineering and rehabilitation, 10(1), 111.

[39] Wijaya, Sastra Kusuina, Cholid Badri, Jusuf Misbach, Tresna Priyana Soemardi, and V. Sutanno. "Electroencephalography (EEG) for detecting acute ischemic stroke." In *2015 4th International Conference on Instrumentation, Communications, Information Technology, and Biomedical Engineering (ICICI-BME)*, pp. 42-48. IEEE, 2015.

[40] Li, X., Chen, X., Yan, Y., Wei, W., & Wang, Z. (2014). Classification of EEG signals using a multiple kernel learning support vector machine. *Sensors*, *14*(7), 12784-12802.

[41] Liu, Ye, Hao Zhang, Min Chen, and Liqing Zhang. "A boosting-based spatial-spectral model for stroke patients' EEG analysis in rehabilitation training." *IEEE Transactions on Neural Systems and Rehabilitation Engineering* 24, no. 1 (2016): 169-179.

[42] Kim, S., Lee, J., & Kim, L. (2016, February). ERS differences between stroke patients and healthy controls after hand movement. In *2016 4th International Winter Conference on Brain-Computer Interface (BCI)* (pp. 1-3). IEEE.https://ieeexplore.ieee.org/document/7457455

[43] Zhang, H., Liu, Y., Liang, J., Cao, J., & Zhang, L. (2013, July). Gaussian mixture modeling in stroke patients' rehabilitation EEG data analysis. In *2013 35th Annual International Conference of the IEEE Engineering in Medicine and Biology Society (EMBC)* (pp. 2208-2211). IEEE.

[44] López-Larraz, Eduardo, Niels Birbaumer, and Ander Ramos-Murguialday. "A hybrid EEG-EMG BMI improves the detection of movement intention in cortical stroke patients

> REPLACE THIS LINE WITH YOUR PAPER IDENTIFICATION NUMBER (DOUBLE-CLICK HERE TO EDIT) <    15